\documentclass[12pt,twoside]{article}

\usepackage{epsf}
\usepackage{rotate}
\usepackage{cite}
\usepackage{times}
\usepackage{fancyheadings}
\advance \headheight by 3.0truept       

\renewcommand{\textfraction}{0}

\newcommand{\lt}{\left}
\newcommand{\rt}{\right}
\newcommand{\ov}{\overline}

\newcommand{\nn}{\nonumber \\}
\newcommand{\no}{\nonumber }
\newcommand{\bbs}{$B_s\!-\!\ov{B}{}_s\,$}
\newcommand{\bbms}{$B_s\!-\!\ov{B}{}_s\,$\ mixing}

\newcommand{\dm}{\ensuremath{\Delta m}}
\newcommand{\dg}{\ensuremath{\Delta \Gamma}}
\newcommand{\eq}[1]{(\ref{#1})}
\newcommand{\guntf}{\ensuremath{ \Gamma  [f,t] }}

\newcommand{\ket}[1]{| \, #1 \, \rangle }
\newcommand{\lqcd}{\Lambda_{\textit{\scriptsize{QCD}}}}
\newcommand{\bea}{\begin{eqnarray}}
\newcommand{\eea}{\end{eqnarray}}
\newlength{\nseparation}
\newenvironment{nfigure}
        {\begin{figure}[tb]\hrule\vspace{\nseparation}\par}
        {\vspace{\nseparation}\par \hrule \end{figure}}

\begin{document}
\thispagestyle{empty}
FERMILAB-Conf-00/231-T\hfill hep-ph/0009203
~\\

\begin{center}
\boldmath
\textbf{\Large
QCD corrections to lifetime differences of 
$B_s$ mesons\footnote{Talk at 
\emph{4th Workshop on Continuous Advances in QCD}, 
12-14 May 2000, Minneapolis, Minnesota, USA}}
\unboldmath
\end{center}
~\\

\begin{center}
Ulrich Nierste\\[3mm]	
Fermi National Accelerator Laboratory\\
Theory Division, MS106\\ 
IL 60510-500 Batavia, USA\\
{\small E-mail: nierste@fnal.gov}
\end{center}

~\\

\begin{center}
\textbf{\large Abstract}
\end{center}
The calculation of QCD corrections to the width difference
\dg\ in the $B_s$-meson system is presented. The next-to-leading order
corrections reduce the dependence on the renormalization scale
significantly and allow for a meaningful use of hadronic matrix
elements from lattice gauge theory.  At present the uncertainty of the
lattice calculations limits the prediction of \dg. The presented work
has been performed in collaboration with Martin Beneke, Gerhard
Buchalla, Christoph Greub and Alexander Lenz.

\section{Introduction}
\renewcommand{\textfraction}{0}
Precision analyses of flavor-changing transitions are of experimental
top priority in the forthcoming years. Decays of $B$ mesons provide an
especially fertile testing ground for various reasons: they allow for
a high precision determination of three of the four parameters
characterizing the Cabibbo-Kobayashi-Maskawa (CKM) matrix\cite{ckm},
including the CP-violating phase $\gamma$.  Since flavor-changing
transitions of $B$ mesons are always suppressed by small CKM elements
and heavy electroweak gauge boson masses, it is well possible that $B$
physics experiments will reveal new physics. The large mass $m_b$ of
the $b$-quark further allows us to control hadronic uncertainties.
Fermilab's CDF, D0\cite{fnal} and the planned BTeV\cite{btev}
experiment prepare a dedicated $B$ physics program. Other studies are
in progress at CLEO, LEP and at HERA-B\cite{herab} or planned for the
future LHCb\cite{lhcb} experiment. While $B$-factories\cite{babar}
only produce $B_d$,$\ov{B}{}_d$ and $B^\pm$ mesons, LEP and the hadron
colliders also provide $B_s$ mesons.  Like their $K$, $D$ and $B_d$
counterparts $B_s$ mesons mix with their antiparticles. Therefore the
two mass eigenstates $B_H$ and $B_L$ (for ``heavy'' and ``light'') are
linear combinations of $B_s$ and $\ov{B}_s$ and differ in their mass
and width.  In the Standard Model \bbms\ is described in the lowest
order by the box diagram depicted in Fig.~\ref{box}.
\begin{nfigure}
\centerline{
\epsfxsize=13pc 
\epsfbox{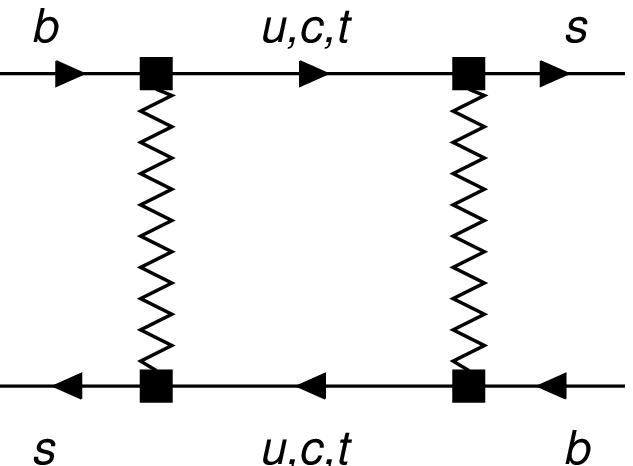}} 
\caption{  \bbms\ in the Standard Model. The zigzag lines represent
$W$-bosons or charged pseudo-goldstone bosons.}\label{box}
\end{nfigure}
The dispersive part of the \bbms\ amplitude is called $M_{12}$. In the
Standard Model it is dominated by box diagrams with internal top
quarks. The absorptive part is denoted by $\Gamma_{12}$ and mainly
stems from box diagrams with light charm
quarks. $\Gamma_{12}$ is generated by decays into final states which are
common to $B_s$ and $\ov{B}_s$. While $M_{12}$ can receive sizable
corrections from new physics, $\Gamma_{12}$ is induced by the
CKM-favored tree-level decay $b \to c \ov{c} s$ and insensitive to new
physics.  Experimentally \bbms\ manifests itself in damped
oscillations between the $B_s$ and $\ov{B}_s$ states which are
governed by $M_{12}-i \Gamma_{12}/2$. We denote the mass and width
differences between $B_H$ and $B_L$ by \bea \dm &=& M_H - M_L \,,
\qquad \dg \; = \; \Gamma_L - \Gamma_H . \no \eea By solving the
eigenvalue problem of $M_{12}-i \Gamma_{12}/2$ one can relate \dm\ and
\dg\ to $M_{12}$ and $\Gamma_{12}$: \bea \dm &=& 2\, |M_{12}|, \qquad
\quad \dg \; =\; 2\, |\Gamma_{12}| \cos \phi, \label{dgsol} \eea where
$\phi$ is defined as \bea \frac{M_{12}}{\Gamma_{12}} = - \lt|
\frac{M_{12}}{\Gamma_{12}} \rt|\, e^{i \phi} \label{defphi}.  \eea
\dm\ equals the \bbs\ oscillation frequency and has not been measured
yet, but we know the lower bound $\dm \geq 14.9\,$ps${}^{-1}$ from LEP
data\cite{os}. It can be shown that this bound implies
$|\Gamma_{12}|/|M_{12}| \ll 0.01 $.  In deriving \eq{dgsol} terms of
order $|\Gamma_{12}/M_{12}|^2$ have been neglected. $\phi$ in
\eq{defphi} is a CP-violating phase, which is tiny in the Standard
Model, so that $\dg_{SM}=2|\Gamma_{12}|$. In the presence of new
physics $\arg M_{12}$ and thereby $\phi$ can assume any value. $\phi$
can be measured from CP-asymmetries, which requires the resolution of
the rapid \bbs\ oscillations and tagging, i.e.\ the discrimination
between $B_s$ and $\ov{B}_s$ mesons at the time $t=0$ of their
production.  From \eq{dgsol} one verifies that a non-vanishing $\phi$
also affects \dg, which can be measured from untagged data samples and
therefore involves better efficiencies than tagged studies. Unlike in
the case of $B_d$ mesons, the Standard Model predicts a sizable width
difference \dg\ in the $B_s$ system, roughly between 5 and 30\% of the
average total width $\Gamma=(\Gamma_L+\Gamma_H)/2$. Now the decay of
an untagged $B_s$ meson into the final state $f$ is in general
governed by two exponentials: 
\bea \guntf &\propto& 
e^{-\Gamma_L t} \lt| \langle f \ket{B_L} \rt|^2 + e^{-\Gamma_H t} \lt|
\langle f \ket{B_H} \rt|^2 . \label{twoex} 
\eea 
If $f$ is a flavor-specific final state like $D_s^- \pi^+$ or $X
\ell^+ \nu$, the coefficients of the two exponentials in \eq{twoex}
are equal. A fit of the corresponding decay distribution to a single
exponential then determines the average width $\Gamma$ up to
corrections of order $(\dg)^2/\Gamma$. In the Standard Model CP
violation in \bbms\ is negligible, so that we can simultaneously
choose $B_L$ and $B_H$ to be CP eigenstates and the $b \to c\ov{c} s$
decay to conserve CP.  Then $B_H$ is CP-odd and cannot decay into a
CP-even double-charm final state $f_{CP+}$ like $(J/\psi
\phi)_{L=0,2}$, where $L$ denotes the quantum number of the orbital
angular momentum. Thus a measurement of the $B_s$ width in $B_s \to
f_{CP+}$ determines $\Gamma_L$. By comparing the two measurements one
finds $\dg/2$. In the presence of a non-zero CP-violating phase $\phi$
this procedure measures\cite{g} \bea \dg \cos\phi &=& \dg_{SM} \cos^2
\phi . \label{dgnp} \eea The extra factor of $\cos \phi$ stems from
the fact that in the presence of CP violation both $B_L$ and $B_H$ can
decay into $f_{CP+}$. CDF will perform this measurement with $B_s \to
D_s^- \pi^+$ and $B_s \to J/\psi \phi $ in Run-II of the
Tevatron\cite{mm}.

\section{QCD effects}\label{sect:qcd}
The \bbms\ amplitude of Fig.~\ref{box} and the $B_s$ decay amplitude
are affected by strong interaction effects. $\dg_{SM}=2 |\Gamma_{12}|$
involves various different mass scales and the the QCD corrections
associated with these scales require different treatments. In the
first step an operator product expansion at the scale $M_W$ is
performed to integrate out the $W$-boson. The Standard Model $b \to
c\ov{c} s$ amplitude is matched to its counterpart in an effective
field theory in which $\Delta B=1$ transitions ($B$ is the bottom
number) are described by four-quark operators. The corresponding
effective hamiltonian reads
\begin{equation}\label{hpeng}
{\cal H}_{eff}=\frac{G_F}{\sqrt{2}}V^*_{cb}V_{cs}
\left(\,\sum^6_{r=1} C_r Q_r + C_8 Q_8\right),
\end{equation}
with the operators
\begin{equation}\label{q1q2}
Q_1= (\bar b_ic_j)_{V-A}(\bar c_js_i)_{V-A}\qquad
Q_2= (\bar b_ic_i)_{V-A}(\bar c_js_j)_{V-A},
\end{equation}
\begin{equation}\label{q3q4}
Q_3= (\bar b_is_i)_{V-A}(\bar q_jq_j)_{V-A}\qquad
Q_4= (\bar b_is_j)_{V-A}(\bar q_jq_i)_{V-A},
\end{equation}
\begin{equation}\label{q5q6}
Q_5= (\bar b_is_i)_{V-A}(\bar q_jq_j)_{V+A}\qquad
Q_6= (\bar b_is_j)_{V-A}(\bar q_jq_i)_{V+A},
\end{equation}
\begin{equation}\label{q8}
Q_8= \frac{g}{8\pi^2}m_b\, 
\bar b_i\sigma^{\mu\nu}(1-\gamma_5)T^a_{ij} s_j\, G^a_{\mu\nu}.
\end{equation}
Here the $i,j$ are colour indices and a summation over $q=u$, $d$,
$s$, $c$, $b$ is implied.  $V\pm A$ refers to
$\gamma^\mu(1\pm\gamma_5)$ and $S-P$ (which we need below) to
$(1-\gamma_5)$. The current-current operators $Q_1$ and $Q_2$ stem
from $W$-boson exchange between the $\ov{b}c$ and $\ov{c} s$
lines. $Q_{3-6}$ are four-quark penguin operators and $Q_8$ is the
chromomagnetic penguin operator. The Wilson coefficients $C_i$ contain
the short-distance physics and are functions of the heavy $W$ and top
quark masses. Since they do not depend on long-distance QCD effect,
they can be calculated in perturbation theory.  $C_{3-6}$ are very
small. The matching calculation determines the $C_i$'s at a high
renormalization scale $\mu={\cal O}(M_W)$. The renormalization group
(RG) evolution of the coefficients down to $\mu ={\cal O}(m_b)$ sums
the large logarithm $\alpha_s \ln (M_W/m_b)$ to all orders in
perturbation theory. The operator product expansion leading to
\eq{hpeng} and the RG improvement amount to a simultaneous expansion
in $m_b^2/M_W^2$, $\alpha_s (M_W)$ and $\alpha_s (m_b)$ of the $b \to
c\ov{c} s$ amplitude. ${\cal H}_{eff}$ in \eq{hpeng} reproduces the
leading term in the power expansion in $m_b^2/M_W^2$.

The second step to predict $\dg_{SM}$ involves an operator product
expansion at the scale $m_b$. The corresponding formalism has been 
formulated long ago by the hosts of this conference\cite{hqe}. 
The starting point for the calculation of the widths $\Gamma_H$ of some
$b$-flavored hadron $H$ is the optical theorem, which relates
$\Gamma_H$ to the absorptive part of the forward scattering amplitude 
of $H$. Neglecting CP violation in the decay amplitude the optical
theorem implies
\begin{eqnarray}
\Gamma_H  
& \propto & 
\mbox{Im}\, \langle H | 
\,i\int d^4x\ T\,{\cal H}_{eff}(x){\cal H}_{eff}(0) |H \rangle .
 \label{opt}
\end{eqnarray}
Now the \emph{Heavy Quark Expansion}\ (HQE)\cite{hqe} is an operator
product expansion of the forward scattering amplitude in \eq{opt}.
Schematically
\begin{eqnarray}
\Gamma_H 
& \propto & G_F^2 \, \sum_j m_b^{8-d_j} \, F_j \, 
                \lt( \mu/m_b  \rt) \, 
        \underbrace{\langle H | {\cal O}_j \lt(\mu \rt) |H \rangle} 
	\label{hqe} \\
&& \hspace{4.5cm} {\cal O} \lt( \lqcd ^{d_j-3}  \rt) \no
\end{eqnarray}  
Here new Wilson coefficients $F_j$ have appeared. They contain the
physics associated with scales above the matching scale $\mu={\cal O}
(m_b)$, at which the HQE is performed. The ${\cal O}_j$'s are local
operators with dimension $d_j \geq 3$. The HQE is a simultaneous expansion
of $\Gamma_H$ in $\lqcd/m_b$ and $\alpha_s (m_b)$. Increasing
powers of $\Lambda_{QCD}/m_b$ correspond to increasing dimensions
$d_j$ of the local operators ${\cal O}_j$. 

To calculate \dg\ from the HQE one must extend the above formalism to
the two state system $(B_s,\ov{B}_s)$:
\begin{eqnarray}
\dg_{SM} & = & 2 |\Gamma_{12}|  
\; =  \; - \frac{1}{M_{B_s}}\,
\mbox{Im}\, \langle\ov{B}_s| 
\,i\int d^4x\ T\,{\cal H}_{eff}(x){\cal H}_{eff}(0)
|B_s\rangle \label{dgopt}
\end{eqnarray} 
The corresponding leading-order diagrams are shown in Fig.~\ref{fig:b2}.
\begin{nfigure}
\centerline{
\epsfxsize=0.8\textwidth 
\epsfbox{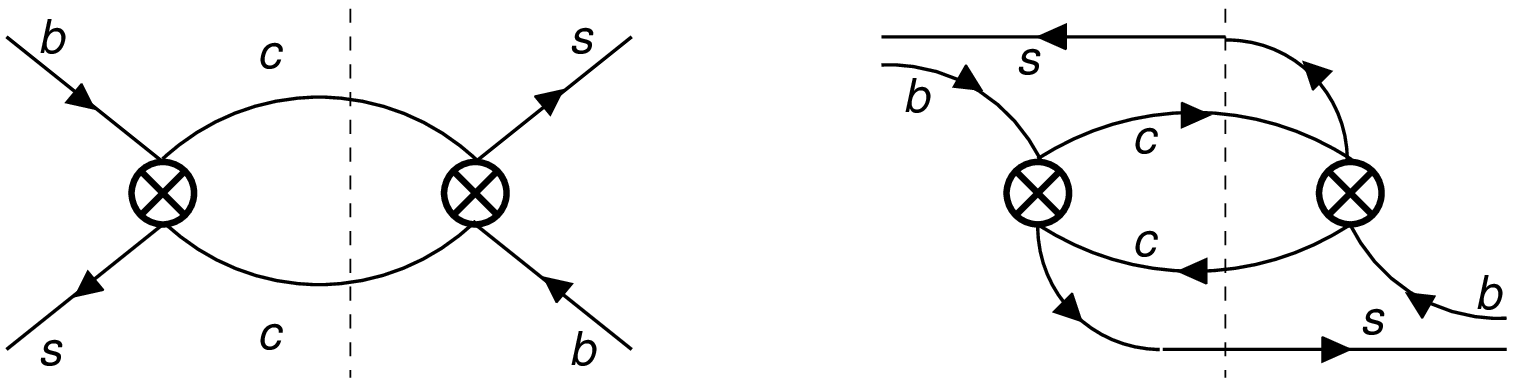}}
\caption{Leading-order diagrams for $\Gamma_{12}$}\label{fig:b2}
\end{nfigure}
\eq{dgopt} is matched to local operators in analogy to \eq{hqe}:
\begin{eqnarray}
\lefteqn{
\hspace{-2ex}
\mbox{Im}\, \langle\ov{B}_s| 
\,i\int d^4x\ T\,{\cal H}_{eff}(x){\cal H}_{eff}(0)
|B_s\rangle} \nn
&=& -\frac{G^2_F m^2_b}{12\pi} \lt| V^*_{cb}V_{cs} \rt|^2
\, \cdot \nn
&&\!\!
\left[ F\lt( \frac{m_c^2}{m_b^2} \rt) 
\langle\ov{B}_s| Q |B_s\rangle + 
F_S \lt( \frac{m_c^2}{m_b^2} \rt)  \langle\ov{B}_s| Q_S |B_s\rangle \right] 
\lt[1+ {\cal O} \lt( \frac{\lqcd}{m_b} \rt) \rt] \!\! .
\label{dghqe}
\end{eqnarray}
The HQE for the $\Delta B=2$  transition in \eq{dgopt} requires 
four-quark operators
involving both the $b$-quark and the $s$-quark field, i.e.\ operators
with dimension six or higher. The two dimension-6 operators appearing in 
\eq{dghqe} are 
\begin{equation}\label{qqs}
Q = (\bar b_is_i)_{V-A}(\bar b_js_j)_{V-A},\qquad
Q_S= (\bar b_is_i)_{S-P}(\bar b_js_j)_{S-P} .
\end{equation}
In the leading order of QCD the RHS of \eq{dghqe} is pictorially
obtained by simply shrinking the $(c,\ov{c})$ loop in
Fig.~\ref{fig:b2} to a point. The Wilson coefficients $F$ and $F_S$
also depend on the charm quark mass $m_c$, which is formally treated
as a hard scale of order $m_b$, since $m_c \gg \lqcd$. Strictly
speaking, the HQE in \eq{dghqe} is an expansion in
$\lqcd/\sqrt{m_b^2-4 m_c^2}$.  For the calculation of $F$ and $F_S$ it
is crucial that these coefficients do not depend on the infrared
structure of the process. In particular they are independent of the
QCD binding forces in the external $B_s$ and $\ov{B}_s$ states in
\eq{dghqe}, so that they can be calculated in perturbation theory at
the parton level. The non-perturbative long-distance QCD effects
completely reside in the hadronic matrix elements of $Q$ and $Q_S$.

The third and final step in the prediction of $\dg_{SM}$ is the
calculation of the hadronic matrix elements with non-perturbative
methods such as lattice gauge theory. It is customary to parametrize
these matrix elements as 
\begin{eqnarray}
\langle\ov{B}_s|Q|B_s\rangle &=& \frac{8}{3}f^2_{B_s}M^2_{B_s} { B} 
\nn
\langle\ov{B}_s|Q_S|B_s\rangle &=& -\frac{5}{3}f^2_{B_s}M^2_{B_s}
\frac{M^2_{B_s}}{(\bar m_b+\bar m_s)^2} { B_S} . \label{me}
\end{eqnarray}
In the so called vacuum insertion approximation $B$ and $B_S$ are
equal to 1. 

\section{Next-to-leading order QCD corrections to \dg}
The discussion of \dg\ in sect.~\ref{sect:qcd} has been restricted to
the leading order (LO) of QCD\cite{LO}. The only QCD effects included
in this order are the leading logarithms $\alpha_s^n \ln^n (M_W/m_b)$,
$n=0,1,2,\ldots$, contained in the $C_j$'s of the effective $\Delta
B=1$ hamiltonian in \eq{hpeng}.  To predict \dg\ with next-to-leading
order (NLO) accuracy one must first include the corrections of order
$\alpha_s^{n+1} \ln^n (M_W/m_b)$, $n=0,1,2,\ldots$, to these
coefficients\cite{bjlw}. Second corrections of order $\alpha_s(m_b)$
must be included in $F$ and $F_S$\cite{bbgln}. This step requires the
inclusion of hard gluon exchange on both sides of \eq{dghqe}. The
corresponding diagrams are depicted in Fig.~\ref{fig:nlo}.
\begin{nfigure}
\centerline{
\epsfysize=0.99\textwidth 
\rotate[r]{\epsfbox{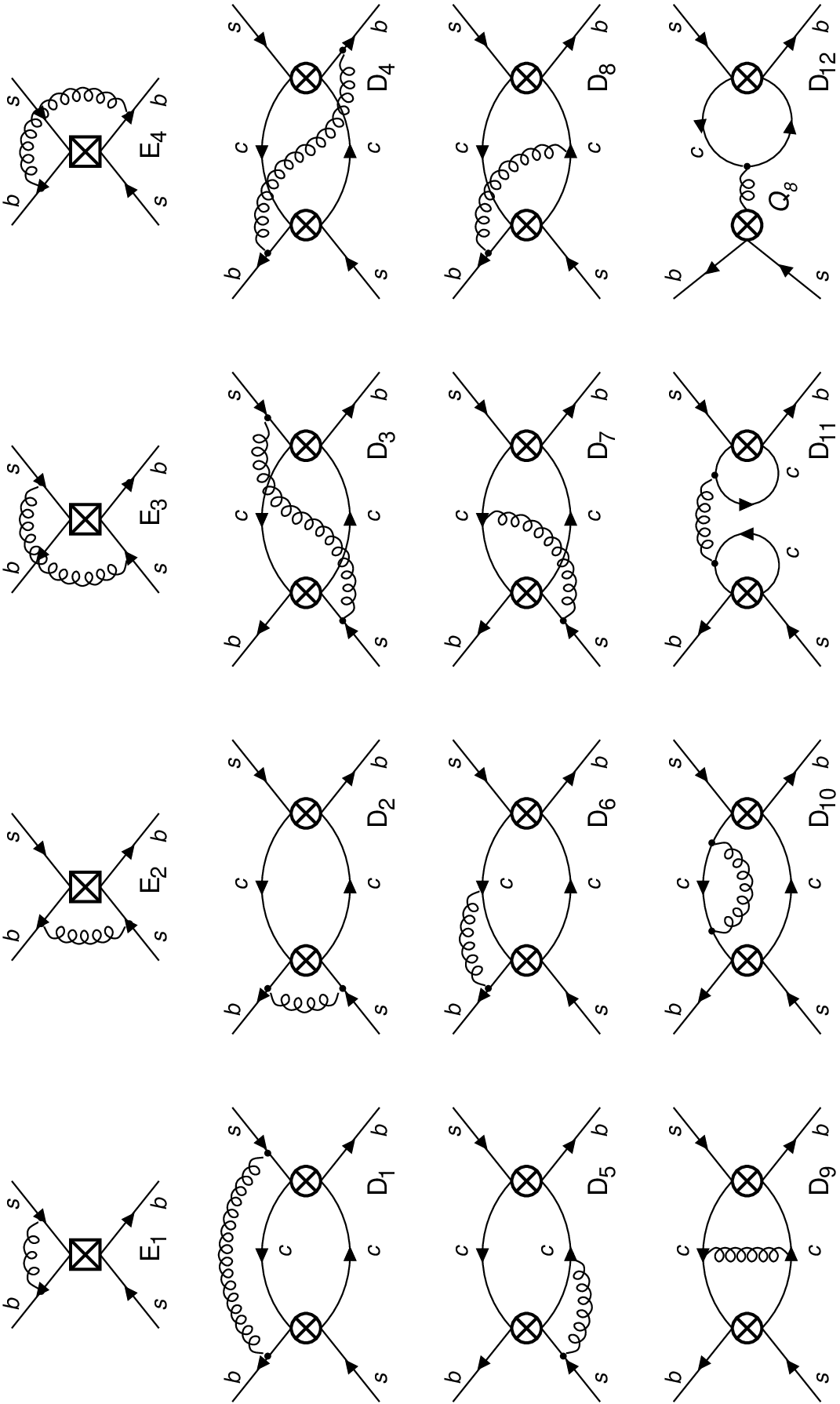}}
}\caption{QCD corrections to \dg. NLO corrections to diagrams with
the penguin operators $Q_{3-6}$ are negligible. }\label{fig:nlo}
\end{nfigure}
The motivations for this cumbersome calculation are
\begin{itemize}
\item[1.] to verify the infrared safety of $F$ and $F_S$,
\item[2.] to allow for an experimental test of the HQE,
\item[3.] a meaningful use of lattice results for 
	hadronic matrix elements,
\item[4.] to reduce the sizable $ \mu$-dependence of the LO,
\item[5.] a consistent use of $ \Lambda_{\ov{\textit{\scriptsize{MS}}}}$,
\item[6.] the large size of QCD corrections, typically of order $30\% $. 
\end{itemize}
We will now comment on these points: A necessary condition for the
validity of any operator product expansion is the disappearance of 
all infrared regulators from the Wilson coefficients. From our
explicit calculation we have verified that this is indeed the case at
order $\alpha_s$. We found IR-singularities to cancel via two
mechanisms:
\begin{itemize}
\item Bloch-Nordsiek cancellations among different cuts of the 
        same diagram,
\item factorization of IR-singularities, which end up in 
         $ \langle \bar B_s | {\cal O}   | B_s \rangle$, 
   $ \langle \bar B_s | {\cal O}_S | B_s \rangle$.
\end{itemize}  
Early critics of the HQE had found power-like infrared divergences in
individual cuts of diagrams. In response the cancellation of these
divergences has been shown\cite{bu}, long ago before we have performed
the full NLO calculation. The second type of IR-cancellations occurs
between the diagrams in the first and second row of
Fig.~\ref{fig:nlo}. Thus when the external meson states in \eq{dghqe}
are replaced by quark states, both sides of the equation are infrared
divergent. Yet the IR-divergences factorize rendering $F$ and $F_S$
infrared safe. Point 2 above addresses the conceptual basis of the
HQE, which is sometimes termed \emph{quark-hadron duality}. It is not
clear, whether the HQE reproduces all QCD effects completely.
Exponential terms like $\exp (-\kappa m_b/\lqcd)$, for example, cannot
be reproduced by a power series.\cite{s} The relevance of such
corrections to the HQE can at present only be addresses
experimentally, by confronting HQE-based predictions with data. The
only QCD informations contained in the LO prediction for \dg\ are the
coefficients of $\alpha_s^n \ln^n M_W$, associated with hard gluon
exchange along the $W$-mediated $b \to c\ov{c} s$ amplitude. The
question of quark-hadron duality, however, has nothing to do with
these logarithmic terms. A meaningful test of this aspect of the HQE
therefore requires a NLO calculation, which includes non-logarithmic
terms of order $\alpha_s$. In view of the success of the HQE in
accurately measured $B$ physics observables it is conceivable that the
uncertainty due to violations of quark-hadron duality is well below
the uncertainty from the non-perturbative calculation of the hadronic
$B$-parameters. At present lattice calculation of $B$ and
$B_s$\cite{hioy} are only possible in the quenched approximation,
neglecting the effect of dynamical fermions.  Unquenched calculations
of $f_{B_s}$ are now available, but still a new subject in the
field\cite{fbs}. The third point in our list above refers to the fact
that QCD predictions obtained on the lattice must be matched to the
continuum. This involves the calculation of the diagrams in the first
row on Fig.~\ref{fig:nlo} in lattice perturbation theory. A meaningful
prediction for \dg\ with a proper cancellation of the renormalization
scale and scheme dependences between $F$,$F_S$ and $B$,$B_S$ then
requires a full NLO calculation. The renormalization scale $\mu$ is an
unphysical parameter and observables do not depend on $\mu$. The
truncation of the perturbation series, however, introduces a
$\mu$-dependence, which diminishes order-by-order in $\alpha_s$.  As
mentioned in point 4, the LO result for \dg\ suffers from a huge scale
dependence, which is substantially reduced in the NLO prediction.
Further a LO calculation cannot use the fundamental QCD scale
parameter $\Lambda_{\ov{\textit{\scriptsize{MS}}}}$\cite{bbdm}, which
is an intrinsic NLO quantity.  Finally, as mentioned in point 6, the
calculated QCD corrections are sizable, of the order of 30\%, and
therefore necessary to keep up with the precision of the forthcoming
experiments.
 
Including corrections of order $\lqcd/m_b$\cite{bbd} to \eq{dghqe}  
we predict\cite{bbgln} 
\begin{eqnarray}\label{dgabc}
\left(\frac{\Delta\Gamma_{SM}}{\Gamma}\right)_{B_s} 
&=& 
\left(\frac{f_{B_s}}{245~{\rm MeV}}\right)^2
\left[0.008\, {B} + 0.204\, {B_S}  - 0.086\right] \no
\end{eqnarray}
with $B$ and $B_S$ defined in the $\ov{\rm MS}$-scheme at $\mu=m_b$. 
With\cite{hioy} 
\begin{eqnarray}
B (\mu= m_b)  &= & 0.80 \pm 0.15, \qquad  
B_S (\mu =m_b) \; = \; 1.19\pm 0.20 \no
\end{eqnarray}
one finds
\begin{eqnarray}
\left(\frac{\Delta\Gamma}{\Gamma}\right)_{B_s}=
\left(\frac{f_{B_s}}{245~{\rm MeV}}\right)^2 \,
\left(0.162 \pm 0.041 \,\pm\, ??? \right) 
. \label{res}
\end{eqnarray}
The questions marks address the unknown error from the quenching
approximation. 

If \dg\ is found below the Standard Model prediction, it will be
interesting to find out, whether this is due to a breakdown of the HQE
prediction in \eq{res} or a new CP-violating phase $\phi$ in
\eq{dgnp}.  To this end we note that one can determine $\phi$ without
using the theory prediction for $\dg_{SM}$, even from untagged data
alone \cite{g,dfn}. Further the HQE prediction for other width
differences, e.g.\ between the $B^+$ and $B_d$ or between the $B_s$
and $B_d$ mesons, involve a similar structure than the prediction for
\dg. The corresponding diagrams are similar to those in
Fig.~\ref{fig:b2} and Fig.~\ref{fig:nlo}, but involve $\Delta B=0$
transitions\cite{hqe,bbd}.  The width difference between $B^+$ and
$B_d$ is insensitive to new physics and therefore directly tests the
HQE and the lattice calculations of hadronic matrix elements. The
small width difference between $B_s$ and $B_d$ is mildly sensitive to
new physics from penguin contributions \cite{kn}.
 
\section*{Acknowledgments}
I thank Misha Voloshin for inviting me to this beautiful workshop and
for the hospitality at his institute. I am grateful for many stimulating
discussions with the participants.

\end{document}